\documentclass[10pt]{iopart}
\usepackage{graphicx}
\usepackage{bm}
\usepackage{iopams} 
\usepackage{datetime}
\usepackage{color}
\usepackage{float}

\begin{document}

\title{The influence of atmosphere on the performance of pure-phase WZ and ZB InAs nanowire transistors}

\author{A~R~Ullah$^{1}$, H~J~Joyce$^{2,3}$, H~H~Tan$^{2}$, C~Jagadish$^{2}$ and A~P~Micolich$^{1}$}
\address{$^{1}$ School of Physics, University of New South Wales, Sydney NSW 2052, Australia}
\address{$^{2}$ Department of Electronic Materials Engineering, Research School of Physics and Engineering, The Australian National University, Canberra ACT 2601, Australia}
\address{$^{3}$ Department of Engineering, University of Cambridge, 9 JJ Thomson Avenue, Cambridge CB3 0FA, United Kingdom}
\ead{adam.micolich@nanoelectronics.physics.unsw.edu.au}

\begin{abstract}
We compare the characteristics of phase-pure MOCVD grown ZB and WZ InAs nanowire transistors in several atmospheres: air, dry pure N$_2$ and O$_2$, and N$_2$ bubbled through liquid H$_2$O and alcohols to identify whether phase-related structural/surface differences affect their response. Both WZ and ZB give poor gate characteristics in dry state. Adsorption of polar species reduces off-current by $2-3$ orders of magnitude, increases on-off ratio and significantly reduces sub-threshold slope. The key difference is the greater sensitivity of WZ to low adsorbate level. We attribute this to facet structure and its influence on the separation between conduction electrons and surface adsorption sites. We highlight the important role adsorbed species play in nanowire device characterisation. WZ is commonly thought superior to ZB in InAs nanowire transistors. We show this is an artefact of the moderate humidity found in ambient laboratory conditions: WZ and ZB perform equally poorly in the dry gas limit yet equally well in the wet gas limit. We also highlight the vital role density-lowering disorder has in improving gate characteristics, be it stacking faults in mixed-phase WZ or surface adsorbates in pure-phase nanowires.
\end{abstract}

\submitto{\NT}
\newline
\noindent Version:~\today~~\currenttime
\maketitle
\ioptwocol

\section{Introduction} One of the most remarkable aspects of III-V nanowires is the ability to access the wurtzite crystal phase~\cite{HirumaJAP95}; bulk non-nitride III-V semiconductors normally only present as zinc blende. Controlled axial changes between the two crystal structures can be achieved by appropriate control over growth conditions~\cite{AlgraNat08}. The two phases differ in bandgap by $\sim 55$~meV for InAs~\cite{ZanolliPRB07}, enabling the production of devices, e.g., quantum dots via a `phase heterostructuring' approach~\cite{AkopianNL10} rather than by compositional heterostructuring. Phase-pure nanowires of substantial length have been grown using Metal-Organic Chemical Vapour Deposition (MOCVD)~\cite{JoyceNL10} and Molecular Beam Epitaxy (MBE)~\cite{PanNL14}. These phase-pure nanowires have since been used to study how crystal phase influences the electronic~\cite{UllahPSSRRL13, FuNL16}, optical~\cite{SunNL12} and thermal properties~\cite{ZhouPRB11}.

Another key feature of semiconductor nanowires is their high surface-to-volume ratio, which makes conduction highly sensitive to surface adsorbates. This has resulted in significant research towards nanowires as electronic gas sensors, e.g., ZnO and In$_2$O$_3$ nanowires have been used for sensing O$_2$, NO$_2$, and NH$_3$~\cite{LiAPL03, FanAPL04}. The surface states in InAs pin the surface Fermi energy at the conduction band leading to a two-dimensional electron gas immediately beneath the nanowire surface. As a result, InAs nanowire conductivity should also be useful for sensing. Several groups have demonstrated gas/biochemical sensing for InAs nanowires with mixed/unknown phase structure~\cite{DuNL09, OffermansNL10, ZhangSensActB14}, but as yet there has been no reports on how gas sensing varies between pure-phase wurtzite (WZ) and pure-phase zinc blende (ZB) nanowires. There is good reason to investigate and expect differences -- the surface facet structure differs significantly between the two phases~\cite{JoyceNL10} and our earlier work revealed differences in electronic properties that were possibly associated with that difference in surface facet structure~\cite{UllahPSSRRL13}. In this paper we present a study of the comparative gas sensitivity of pure-phase WZ and ZB nanowires. We show that WZ shows greater sensitivity in the low adsorbate level limit, presumably due to surface-facet-related differences between the two phases. We also discuss the important bearing this has on interpretations of gate performance characteristics, and transport studies more broadly, in InAs nanowires.

\section{Device Fabrication and Electrical Measurement} InAs$(\bar{1}\bar{1}\bar{1})$B substrates were treated with poly-L-Lysine (Sigma, $0.1\%$ in H$_2$O), followed by solution deposition of $50$~nm diameter Au nanoparticles (Ted Pella) and a $600^{\circ}$C anneal in AsH$_3$ to desorb surface contaminants. Phase-pure InAs nanowires were grown using horizontal flow MOVPE with trimethylindium (TMI) and AsH$_3$ (hydrides from Air Products at $>99.9999\%$ and metalorganics from SAFC Hitech at epipure grade). Growth was performed at $400^{\circ}$C ($500^{\circ}$C) for ZB (WZ) with TMI flow rate $1.2 \times 10^{-5}$~mol/min and AsH$_3$ flow rates $5.5 \times 10^{-4}$~mol/min ($3.5 \times 10^{-5}$~mol/min). The corresponding III/V ratios were $46$ ($2.9$). This resulted in ZB (WZ) nanowires with their axis along the $[\bar{1}\bar{1}\bar{1}]$ ($[000\bar{1}]$) direction. Further details can be found in Ref.~\cite{JoyceNL10}. The nanowires are exposed to air after growth, during processing and measurement and thus will have a thin, self-limiting layer of native oxide on their surface, except at the contacts where it is removed by the passivation treatment prior to metallisation (see below). High-resolution transmission electron microscopy (HRTEM) images and structural characterisations of the same nanowires studied here appear in prior publications~\cite{JoyceNL10, UllahPSSRRL13}.

\begin{figure} \centering \includegraphics[width=\columnwidth]{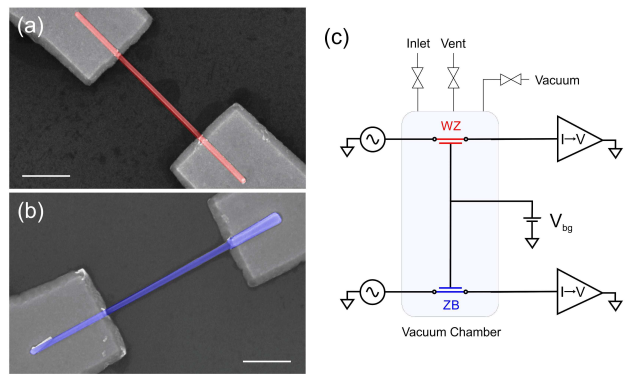} \caption{(a/b) Scanning electron micrographs for typical (a) WZ and (b) ZB InAs nanowire transistors. The scale bars represent $1~\mu$m. (c) Schematic of our measurement apparatus, which enables WZ and ZB devices to be simultaneously measured under common atmospheric conditions. \label{fig:1}} \end{figure}

Nanowires were transferred to a device substrate using dry laboratory wipe. The substrate consists of degenerately doped Si (SVMI), capped with $100$~nm SiO$_2$ and $10$~nm HfO$_2$  by atomic-layer deposition, and featuring $5$~nm Ti/$100$~nm Au patterned interconnects. WZ and ZB nanowires were deposited on separate interconnect fields separated by $3.5$~mm on a single substrate to ensure the processing steps are identical for both phases. Ohmic contacts to the nanowires were defined by electron beam lithography and thermal evaporation of $25$~nm Ni/$75$~nm Au (ESPI Metals 99.99+\%). We passivated the exposed nanowire contact surface immediately prior to metallisation by immersion in (NH$_4$)$_2$S$_x$ solution (Aldrich). The ZB and WZ device fields were separated by cleavage after contact deposition and packaged separately in 20-pin ceramic chip carriers (Spectrum LCC-20). Each chip contains up to $24$ individual WZ or ZB nanowire transistors, depending on yield, with some fraction of those available for measurement due to constraints on the number of wires available. Multiple devices have been measured, both on common chips and separate chips, to confirm repeatability of the results. Scanning electron micrographs of completed WZ and ZB devices are shown in Figs.~\ref{fig:1}(a) and (b), respectively. The WZ nanowires have negligible taper with diameter $70 < d < 100$~nm, whereas the ZB nanowires show some tapering due to radial overgrowth~\cite{JoyceNL10} with diameter ranging from $100$~nm at the narrow end to $300$~nm at the wide end. Scanning electron microscopy was used to determine channel dimensions after electrical measurements were complete to avoid electron-beam induced damage affecting the data. Further details of the growth, structural characterisation and device fabrication can be found in our earlier work~\cite{JoyceNL10, UllahPSSRRL13}.

\begin{figure} \centering \includegraphics[width=\columnwidth]{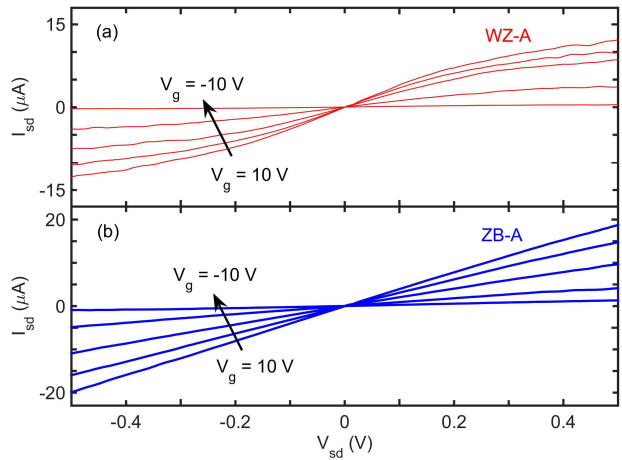} \caption{Plot of $I_{sd}$ vs $V_{sd}$ source-drain characteristics for phase-pure (a) WZ and (b) ZB nanowire transistors measured under dry pure N$_2$. Characteristics are presented for $V_{bg} = +10, +5, 0, -5$ and $-10$~V. \label{fig:2}} \end{figure}

Electrical measurements were performed at room temperature in a specially constructed atmospheric control chamber that doubled as a Faraday cage. The chamber had two sockets for chip packages installed so that a ZB device and a WZ device could be measured together under identical conditions (see Fig.~1(c)). The control chamber was connected to a rotary vacuum pump and gas inlet and outlet lines. For measurements in air or He, the chamber was filled to $1$~atm and left isolated. The He was supplied as warm dry vapour obtained from a liquid helium dewar for convenience. Dry pure N$_2$ and O$_2$ were supplied continuously with flow rate $0.25$~L/min from cylinders. Methanol (MeOH), ethanol (EtOH), 2-propanol (2-PrOH) and H$_2$O vapour were supplied by bubbling dry pure N$_2$ through a glass bubbler filled with the corresponding liquid (all alcohols from Aldrich at reagent grade; water is de-ionized by millipore system). These measurements are also performed under continuous flow. Measurements under air were obtained using room atmosphere with a relative humidity of approximately $55\%$. Measurements for N$_2$ bubbled through H$_2$O result in relative humidity of approximately $85\%$. The MeOH, EtOH and 2-PrOH fractions in N$_2$ are estimated to be approximately $250$~ppm, $130$~ppm and $180$~ppm respectively, with approximately $10\%$ uncertainty. These estimates were obtained by measuring the decrease in liquid volume in the bubbler during the known time period that gas is bubbled through and combining with the N$_{2}$ flow rate of $0.25$~L/min.

\begin{figure} \centering \includegraphics[width=\columnwidth]{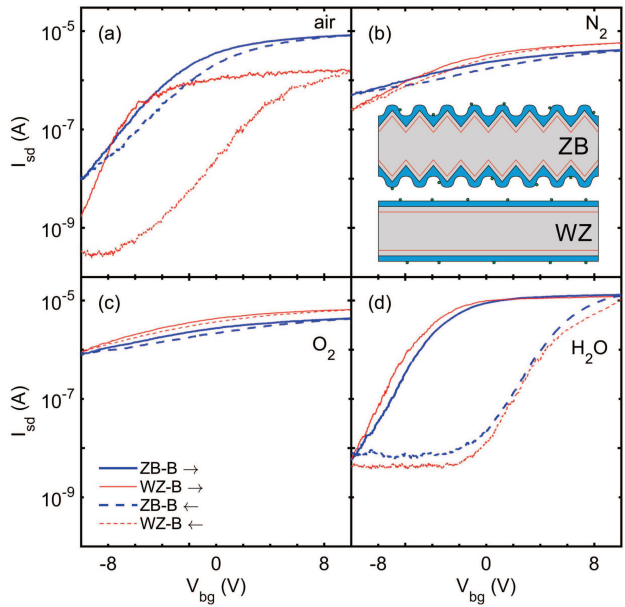} \caption{Plot of $I_{sd}$ vs $V_{bg}$ transfer characteristics for exposure to (a) air, (b) N$_2$, (c) O$_2$ and (d) H$_2$O. Data for ZB is shown in blue and WZ in red, with sweeps from negative to positive $V_{bg}$ shown as solid lines and sweeps from positive to negative $V_{bg}$ shown as dashed lines since the hysteresis is significant. $V_{bg}$ is swept at $1.3$~V/min and $V_{sd} = 100$~mV in all cases. Inset to (b) is a schematic illustrating our point regarding nanowire (grey) facets, oxide (blue) and the separation between surface adsorbed species (green dots) and the sub-surface two-dimensional electron gas (red) in InAs as discussed in \S3.2. The ZB case is deliberately exaggerated graphically for clarity; the true extent of this effect geometrically may be less than drawn.~\label{fig:3}} \end{figure}

The back-gate voltage $V_{bg}$ was supplied by a Keithley K2410 source measure unit. The channel current measurements were performed using dc methods for the source-drain characteristics and ac lock-in methods for the gate (transfer) characteristics. The dc measurements have a variable source-drain voltage $V_{sd}$ supplied by a K2410 with the source-drain current $I_{sd}$ measured using a Keithley K6517a electrometer. In the ac measurements, $I_{sd}$ was measured using a Stanford SR830 lock-in amplifier in series with a Femto DLPCA-200 current pre-amplifier, with a constant $V_{sd} = 100$~mV(rms) at $73$~Hz.

\begin{figure} \centering \includegraphics[width=\columnwidth]{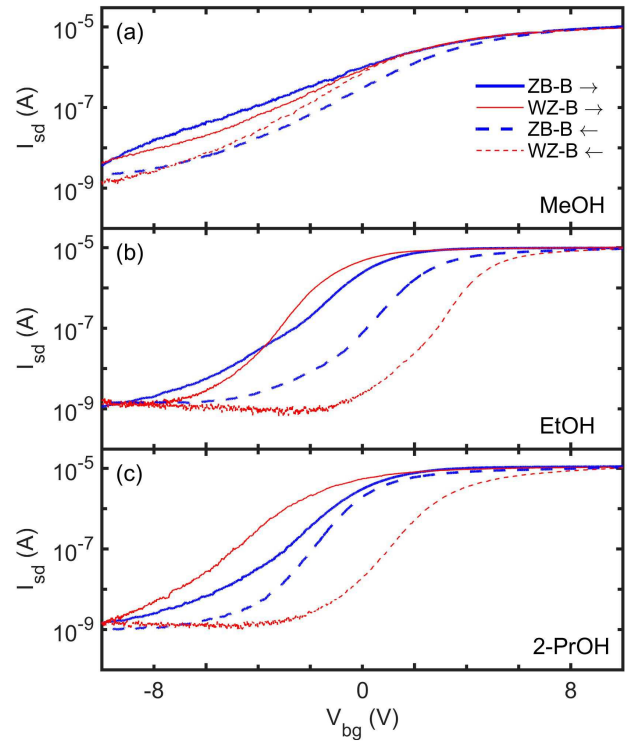} \caption{Plot of $I_{sd}$ vs $V_{bg}$ transfer characteristics for exposure to (a) MeOH, (b) EtOH and (c) 2-PrOH. Data for ZB is shown in blue and WZ in red, with sweeps from negative to positive $V_{bg}$ shown as solid lines and sweeps from positive to negative $V_{bg}$ shown as dashed lines since the hysteresis is significant. $V_{bg}$ is swept at $1.3$~V/min and $V_{sd} = 100$~mV in all cases.~\label{fig:4}} \end{figure}

The typical measurement protocol applied is as follows. The devices are initially `reset' by baking at $50^{\circ}$C for $5$~min on a hotplate to drive off adsorbed surface species. This temperature is deliberately kept low to avoid altering the contact resistance. Devices are then added to the chamber and initially characterised in air to obtain a starting state for the device. Any dry gas characterisation, typically N$_2$, is then performed. Vapour characterisation is performed after this running from most volatile to least volatile. Water is always the last vapour used as it is the most difficult to remove. When switching between different vapour compositions, the devices are kept under flowing dry pure N$_2$ overnight, and we check that the electrical characteristics return to the pure N$_2$ state before admitting the new vapour. Repeated cycles of measurement under a given vapour and purge with pure dry N$_2$ were used to establish repeatability. To confirm reproducibility we repeated the experiments, with the devices reset between experiments using the bake procedure described above.

\begin{figure} \centering \includegraphics[width=\columnwidth]{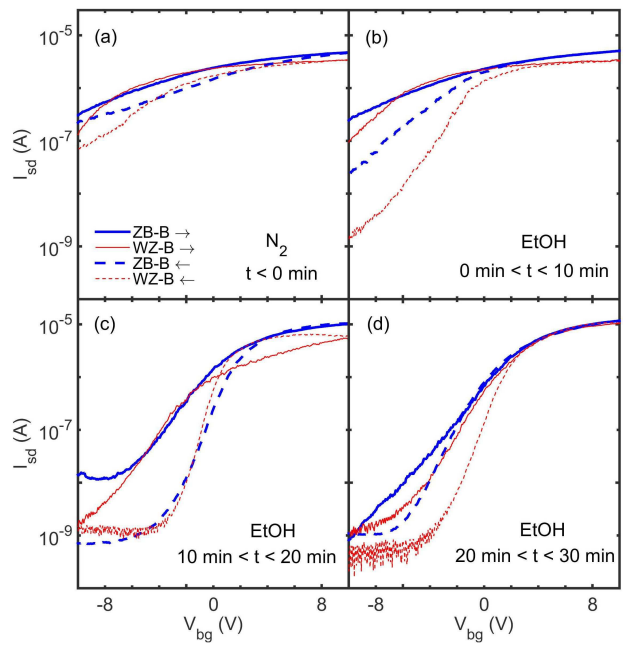} \caption{Plot of $I_{sd}$ vs $V_{bg}$ for (a) exposure to dry pure N$_2$ prior to addition of EtOH at time $t = 0$, and (b-d) with N$_2$ bubbled through EtOH, with the data in (b) obtained during the first 10 minutes, (c) during second 10 minutes and (d) during third 10 minutes of EtOH exposure. In each case $V_{bg}$ is swept at $4$~V/min with characteristics for WZ and ZB obtained simultaneously by independent measurement circuits.~\label{fig:5}} \end{figure}

\section{Results} \subsection{Source-Drain Characteristics} Figure~\ref{fig:2} shows typical source-drain characteristics for the WZ and ZB nanowire transistors used in this work. The WZ devices show saturation for $V_{sd} < 100$~mV whereas the ZB devices show linear behaviour over the entire $V_{sd} = \pm0.5$~V range. A lower saturation voltage $V_{sat}$ for phase-pure WZ nanowires was also found for nanowires grown by MBE~\cite{FuNL16}, and may have two contributing factors. Firstly, there is the smaller diameter $d$ of the WZ nanowires due to the lack of radial overgrowth, since $V_{sat} \propto n{\cdot}{d}^2$, where $n$ is the electron density~\cite{SzeBook06}. It may also arise from the differing $n$ and electron mobility found between WZ and ZB nanowires~\cite{UllahPSSRRL13}. To avoid saturation from affecting the gas sensing data below, we deliberately keep $V_{sd} = 100$~mV~$\leq~V_{sat}$ for all measurements that follow.

\subsection{Effect of Atmospheric Composition on Gate Characteristics}

Figures~3 and 4 show semilogarithmic transfer ($I_{sd}$ vs $V_{bg}$) characteristics for a single ZB device (blue traces) and a single WZ device (red traces) under seven different atmospheres. Gate hysteresis is a significant feature of this data. Thus in each case traces from negative to positive (solid lines) and positive to negative (dashed lines) $V_{bg}$ are presented. Figure~3 focusses on air and key components thereof. Figure~4 focusses on exposure to alcohols, following Ref.~\cite{DuNL09}. The same data plotted on a linear $I_{sd}$ axis is presented in Supplementary Figure~S1 for completeness. The transfer characteristics for vacuum and He are presented in Supplementary Figure~S2 as control studies. In Table~1 we present three key gate characteristic parameters: threshold voltage $V_{th}$, sub-threshold slope $S$ and on-off ratio for the data in Fig.~3/4, along with some inferred quantities that we discuss later.

The most significant aspect of Figs.~3/4 is the marked difference between the moderately-dry air atmosphere, dry N$_2$ atmosphere, wet N$_2$ atmosphere (bubbled through H$_2$O) and N$_2$ plus alcohol atmospheres. The trace in Fig.~\ref{fig:3}(a) was obtained after the `reset' bake was performed and the device was left mounted in the sealed air-filled chamber for $24$~hours to stabilise. Both WZ and ZB devices show gate threshold voltage $V_{th} \sim -10$~V. The off-current for the ZB device is particularly poor, and substantially worse than the WZ device, consistent with earlier work~\cite{DayehAFM09, UllahPSSRRL13}. The other key difference is the substantially enhanced hysteresis for WZ. The key influence here is H$_2$O, as we now demonstrate via the remaining data in Fig.~3. Air has composition $78\%$ N$_2$, $21\%$ O$_2$ plus trace gases. For dry air the two main trace gases are Ar ($1\%$) and CO$_2$  ($400$ ppm). We assume the latter is negligible. The former is inert and should give a response matching the He data in Supplementary Figure~2(c). For non-dry air, H$_2$O is a significant and highly variable component ranging from ppm to $\sim3\%$. The data in Fig.~3(a) at $\sim55\%$ relative humidity corresponds to a water fraction of $17,550$~ppm. We can take the water fraction to zero by filling the chamber with pure N$_2$, which gives the data in Fig.~3(b). Here the characteristics for WZ and ZB are notably similar with very low on-off ratio and minimal hysteresis. The characteristics under pure dry O$_2$ in Fig.~3(c) closely match those for pure N$_2$ demonstrating that oxygen is not responsible for the difference between WZ and ZB in Fig.~3(a). Figure~3(d) shows data obtained under continuous flow of N$_2$ bubbled through H$_2$O. This produces a higher relative humidity than air ($\sim85\%$) corresponding to a water fraction of $27,500$~ppm. Here we find the gate characteristics for both WZ and ZB become strongly hysteretic with more positive $V_{th}$, sharper sub-threshold slope and significantly improved on-off ratio. Evidently, both WZ and ZB respond to adsorbed water according to Fig.~3(d), but Fig.~3(a) demonstrates WZ is more sensitive to low levels of adsorbed water than ZB. We will explain this behaviour in terms of structural differences between WZ and ZB in Section~4.

If we now switch back to pure dry N$_2$ the gate characteristics return to those in Fig.~3(b). This confirms H$_2$O produces the effect we see in Figs.~3(a/d) and that it is reversible, albeit after long exposure to dry N$_2$ (at least overnight) since H$_2$O is very slow to desorb. Exposure to alcohol vapour (see Fig.~4) gives a similar outcome, as expected since H$_2$O, MeOH, EtOH and 2-PrOH are all polar. Adsorbates, e.g., H$_2$O, produce similar effects in devices based on carbon nanotubes~\cite{KimNL03}, graphene~\cite{WangACSNano10} and MoS$_2$~\cite{LateACSNano12}, where the conduction channel is in close proximity to a highly exposed surface. Comparing Fig.~3(d) and 4(a-c), we find the strongest effect for H$_2$O and 2-PrOH, with MeOH closest to dry gas. The effect we see in Figs.~3/4 scales with adsorbate volatility suggesting the strength of the adsorbate-surface interaction is key. The electronic mechanism is likely surface charge transfer as addressed in earlier work~\cite{DuNL09}; we return to this aspect in Section~4.

In Fig.~5 we look at the temporal evolution with EtOH exposure as further evidence for the role of adsorbate-surface interaction on the gate characteristics. We focus on EtOH here as it is intermediate in volatility and thus easiest to discern an effect at the timescales enforced by the period needed to sweep through the full gate characteristic. Figure~5(a) presents the gate characteristics immediately prior to exposure for comparison. As noted earlier, the off-current for both WZ and ZB is high under dry N$_2$ exposure. At time $t~=~0$ we start passing the N$_2$ gas via the EtOH bubbler and record the characteristics -- these characteristics necessarily evolve during the measurement but subsequent `snapshots' demonstrate the effect. The WZ device clearly shows a stronger response to adsorbate at early stage exposure (Fig.~5(b)). The ZB device catches up quickly, as Fig.~5(c/d) show. Equilibrium is reached after approximately 20~min, i.e., traces obtained for $t~>~30$~min match those shown in Fig.~5(d). The transition we show here is reversible; resuming dry pure N$_2$ flow gives a rapid return (approx. 2~hrs) to the characteristics shown in Fig.~5(a). EtOH is significantly better than H$_2$O in this regard, the latter needs at least $12$~hours in dry, pure N$_2$ to return to starting characteristics like those in Fig.~5(a) or 3(b). This is not surprising, it is well known that H$_2$O is particularly difficult to desorb from surfaces without elevated temperatures.

\section{Discussion}
At this point we consider the WZ and ZB data in Figs.~3-5 in terms of structurally-derived differences between the two phases. The WZ nanowires clearly show the greatest sensitivity to adsorbate, as evident for air in Fig.~3(a) and EtOH in Fig.~5(b). This cannot be due to absolute nanowire surface area; the ZB nanowires have greater facetting~\cite{CaroffNatNano09,JoyceNL10} and a larger diameter due to radial overgrowth/tapering (see Fig.~1(a/b)). These combine to give ZB a far greater absolute NW surface area. A possible explanation might relate to differences in the separation between surface adsorbates and conduction electrons due to differences in surface facet structure and the native oxidation of those facets, as illustrated in the schematic inset to Fig.~3(b). Conduction in InAs nanowires is largely due to a near surface two-dimensional electron gas (2DEG), shown in red in the schematic to Fig.~3(b). It is formed by surface-state-induced band-bending, which pins the surface Fermi energy near the conduction band edge~\cite{HeedtNanoscale15}. The exact charge distribution in this 2DEG is influenced by many factors including symmetry, electron-electron repulsion, quantum confinement, impurities both intentional (i.e., doping) and unintentional in the nanowire, surface potential variations, etc.~\cite{HeedtNanoscale15,WeisNanotech14}. Crucially, because the 2DEG is highly exposed to the surface, the charge distribution and thereby conduction can also be influenced by charged species adsorbed to the outside surface of the nanowire. A key structural difference between WZ and ZB nanowires is the crystal facets exposed at the nanowire surface. Most significantly for this work, pure ZB InAs nanowires have a more highly-facetted external surface than pure WZ InAs nanowires~\cite{JoyceNL10}. This has a flow-on effect to the native oxide that grows on any semiconductor surface when it is exposed to air post-growth. The oxide is not crystalline, and as illustrated in the inset to Fig.~3(b), tends to be thicker around sharper changes in surface orientation, e.g., at the valleys and peaks where adjacent facets meet. This is borne out in the high-resolution transmission electron microscope (HRTEM) studies for these specific nanowires, as previously published in~\cite{JoyceNL10,UllahPSSRRL13}. There is considerably less oxide thickness variation for WZ~\cite{JoyceNL10,UllahPSSRRL13} and we believe this is the origin for the greater sensitivity of WZ over ZB for H$_2$O adsorption. In the low adsorbate limit, any species absorbing to the WZ nanowire surface will have a fixed separation from the 2DEG (see Fig.~3(b) inset). In contrast, for ZB the separation depends where the adsorbate lands -- near the middle of a facet the separation will be comparable to that for WZ, but near a facet edge the separation could be much higher due to the increased local oxide thickness. This means some proportion of absorbates for ZB will have a reduced effect leading to the observation of stronger water response for WZ in Fig.~3(a). At higher adsorbate levels there is enough coverage for both WZ and ZB to show a response as in Fig.~3(d). The schematic in Fig.~3(b) is drawn to emphasise the mechanism we propose; in reality the extent may be less than shown but the mechanism would nonetheless still produce what we see in Figs.~3-5. It would be interesting for this to be modelled computationally to better understand the true extent of this effect and we encourage theoretical studies in this direction. The way that facetting affects the comparative surface-to-volume ratio between ZB and WZ should also be accounted for as part of this.

Differences in gate hysteresis for WZ and ZB in Figs.~3-5 may also arise from two facet-related effects. The first is the proximity of adsorbates to the 2DEG or underlying surface states at any given point due to oxide thickness variation, as described above. The second is the nature of the intermediate native oxide, which varies from poly-crystalline to amorphous across the nanowire surface~\cite{JoyceNL10}, thereby influencing any charge transfer that takes place across it.~\cite{DuNL09} The varying nature of the semiconductor surface states for facets with differing Miller indices may also play a role. The combined effect of these explains why the relative sizes of the hysteresis loops vary from device-to-device and even measurement-to-measurement (see Supplementary Fig.~3) despite us consistently finding that WZ has the more sensitive response as discussed earlier.

A notable outcome in our data is the sensitivity of threshold voltage, sub-threshold slope and on-off ratio to adsorbed H$_{2}$O for both WZ and ZB. Prior to our work on comparison of pure-WZ and pure-ZB InAs nanowires the prevailing understanding was that the gate performance for ZB nanowires was universally poor and the superior performance for WZ nanowires arose from axial interface polarization charge effects at stacking faults in these WZ nanowires~\cite{DayehAFM09}. These polarization charge effects have been studied with more depth very recently by Chen {\it et al.}~\cite{ChenNL17}. In our earlier work, we showed that WZ nanowires give good gate performance even in the absence of stacking faults and interface polarization charge effects~\cite{UllahPSSRRL13}. Our ZB nanowires also gave strong performance characteristics~\cite{UllahPSSRRL13}, in spite of earlier findings~\cite{DayehAFM09}. The results we present here provide a better understanding of how these differences in performance arise, and force us to caveat those earlier conclusions in Ref.~\cite{UllahPSSRRL13} slightly. The performance of InAs nanowire transistors is poor for both pure-WZ and pure-ZB if there are no surface adsorbed polar species present, i.e., the system has been held under dry-state conditions such that full desorption occurs. The greater sensitivity of WZ to adsorbed water means that under low to moderate humidity conditions, the WZ nanowires appear to be superior to ZB explaining our results in Fig.~3(a). The adsorbed H$_2$O essentially acts as a surface dopant that reduces the 2DEG density such that pinch-off happens more effectively -- a similar argument is made for the action of the interface polarization charge in terms of aiding pinch-off~\cite{DayehAFM09}. The improved ZB characteristics in Ullah {\it et al.}~\cite{UllahPSSRRL13} over those in Dayeh {\it et al.} are likely just the result of slightly higher adsorbed H$_2$O arising from higher relative humidity of the measuring atmosphere. In the limit of high hydration, as we produce in Fig.~3(d) by running N$_2$ through a water bubbler, we get strong gate performance for both morphologies. One last aspect to note, it may seem that our earlier He data in Ref.~\cite{UllahPSSRRL13} is inconsistent with the above (and data in Supplementary Figure~2(c)). For the case in Ref.~\cite{UllahPSSRRL13} the He atmosphere was obtained by direct immersion of the device into the boil-off gas above the liquid in a helium dewar. This cold gas would almost immediately freeze any adsorbed H$_2$O into place. The He gas in Supplementary Fig.~2(c) was a small volume taken off the top of the dewar where $T \sim 260$~K via a $2$~m long rubber hose, which allowed the gas to equilibrate to room temperature. It is necessarily dry, since any stray H$_2$O in the dewar will be frozen out, and thus it results in the poor gate performance for both WZ and ZB apparent in Supplementary Fig.~2(c).

To close the discussion, in Table~1 we present the corresponding carrier concentration $n$ and field-effect mobility $\mu_{FE}$, along with the threshold voltage $V_{th}$, sub-threshold slope $S$ and on-off ratio values for the data in Figs.~3/4. We obtain the carrier concentration as $n = \frac{C}{e\pi r^2L} \frac{I_{sd}}{g_m}$ from the measured channel current $I_{sd}$ above threshold and the transconductance $g_{m} = \partial~I_{sd}/\partial~V_{bg}$. Here $r$ is the average nanowire radius, $L$ is the channel length, and the capacitance $C = (2\pi\epsilon~L)/ln(\frac{h + r + \sqrt{(h+r)^2 - r^2}}{r})$ via the cylinder-on-plane model~\cite{MartelAPL98}. The relative permittivity of the SiO$_2$/HfO$_2$ dielectric $\epsilon = 4.22\epsilon_{0}$ with SiO$_2$/HfO$_2$ layer thickness $h = 110$~nm. Lastly, the field-effect mobility follows as $\mu_{FE} = g_{m}L^{2}/CV_{sd}$.

\begin{table*}
    \centering
    \begin{tabular}{c c c c c c c c c c c}
    \hline\noalign{\smallskip}
    atmosphere & \multicolumn{2}{c} {$V_{th}$ (V)} & \multicolumn{2}{c} {$S$ (mV/dec)} & \multicolumn{2}{c} {on/off (dec)} & \multicolumn{2}{c} {n ($\times 10^{17}$~cm$^{-3}$)} & \multicolumn{2}{c} {$\mu_{FE}$ (cm$^2$/Vs)}
    \\
    {} & WZ & ZB & WZ & ZB & WZ & ZB & WZ & ZB & WZ & ZB \\
    \hline
    Air & -7.4/2.5 & -4.5/-2.9 & 2320 & 3330 & 3.4 & 3.0 & 4.7 & 1.9 & 340 & 1400 \\
    N$_2$ & -8.0/-6.2 & -10.3/-7.5 & 7600 & 15620 & 1.4 & 1.0 & 6.1 & 2.9 & 370 & 670 \\
    O$_2$ & -12.8/-12.6 & -12.3/-12.0 & 12590 & 19830 & 0.9 & 0.8 & 12.9 & 4.4 & 690 & 410 \\
    H$_2$O & -5.6/6.0 & -4.8/4.4 & 1980 & 2090 & 3.5 & 3.3 & 3.8 & 1.2 & 4780 & 3970 \\
    MeOH & 0.0/-0.2 & -0.1/1.3 & 3090 & 3750 & 3.7 & 3.7 & 5.7 & 1.9 & 2260 & 2250 \\
    EtOH & -2.2/3.6 & -1.0/1.1 & 1100 & 1520 & 3.9 & 3.9 & 2.1 & 0.7 & 4810 & 4490 \\
    2-PrOH & -3.2/2.0 & -1.3/-0.9 & 1560 & 1500 & 3.9 & 4.0 & 3.3 & 0.8 & 3300 & 4220 \\
    \hline
    \end{tabular}
    \caption{Table 1: Threshold voltage $V_{th}$, sub-threshold slope $S$, on-off ratio, carrier concentration $n$
    and field-effect mobility $\mu_{FE}$ for pure-WZ and pure-ZB InAs nanowire transistors under various
    atmospheres. Quantities extracted using data in Figures~3/4.~\label{tab:1}}
\end{table*}

Polar adsorbates significantly decrease $n$ producing a corresponding increase in $\mu_{FE}$ by a factor of the order of $3-10$ for our devices. This is consistent with previous experimental work~\cite{DuNL09,ZhangSensActB14} and explained by Du {\it et al.}~\cite{DuNL09} as arising from the bulk mobility dominating over the surface mobility because the adsorbate reduces $n$ by direct capture of charge carriers. This should produce a corresponding positive shift in $V_{th}$, which we also observe. We find WZ has comparable $\mu_{FE}$ to ZB despite the higher $n$. This points to a higher intrinsic mobility for WZ, consistent with our earlier work~\cite{UllahPSSRRL13}. However, the key point in Table~1 is that the gate performance of InAs nanowire transistors is heavily influenced by surface adsorbed species. The effect has sufficient impact that it is worth considering the implications for prior studies where adsorbed species may have been ignored entirely or at least not carefully controlled for. For studies performed under ambient conditions, variations in humidity between devices/measurements may have dominated over other aspects, e.g., intrinsic mobility, impurity density, variations in growth parameters, etc. We will refrain from working back through this literature here, but it should be kept in mind for future interpretation of these works. One last aspect of note: our data in Figs.~3-5 emphasise the importance of disorder in improving the gate performance of InAs nanowires by lowering the electron density $n$. For mixed phase nanowires this disorder is structural; stacking faults produce interface polarization charge that locally lowers $n$ to improve gate performance~\cite{DayehAFM09,ChenNL17}. For pure phase nanowires this disorder needs to be external, in the form of polar adsorbates and the associated carrier trapping that results.

\section{Conclusions}
We compared the electronic characteristics of phase-pure MOCVD grown ZB and WZ InAs nanowire FETs in various atmospheres: air, dry pure N$_2$, dry pure O$_2$ and N$_2$ bubbled through liquid H$_2$O, methanol, ethanol and $2$-propanol. Control studies were also performed in vacuum and dry pure He gas. We aimed to identify whether structural/surface differences between the ZB and WZ phases affect their response to gaseous atmospheres, partly to address a standing question from our earlier work~\cite{UllahPSSRRL13} and partly for interest in potential gas sensing applications~\cite{DuNL09,ZhangSensActB14}.

In vacuum and dry pure gases we find that both WZ and ZB have poor gate characteristics with low on-off ratio, high off-current and high subthreshold swing. Adsorption of polar species to the surface, e.g., H$_2$O, alcohols, provides significant improvement reducing off-current by $2-3$ orders of magnitude, reducing sub-threshold swing and increasing the on-off ratio. Adsorption leads to a decrease in electron density $n$, as measured from gate characteristics, and this is consistent with an observed positive shift in threshold voltage. The reduced $n$ is consistent with the mechanism of direct carrier trapping by the adsorbate, as proposed in earlier work by Du {\it et al.}~\cite{DuNL09}, and this is likely responsible for the strong gate hysteresis we observe for nanowires exposed to polar species.

The key difference between WZ and ZB in terms of gas response is the sensitivity to low levels of adsorbate. The WZ nanowires are significantly more sensitive to adsorbate, which we confirmed for both H$_2$O and EtOH but expect to hold for all cases. We attribute this difference to facet structure, how this affects native oxide thickness and, as a flow-on effect, the separation between surface adsorbates and the conduction channel in the nanowire. ZB has substantially stronger facetting, which we argue leads to slight oxide thickening at the valleys and peaks between facets, supported by past high-resolution TEM studies.~\cite{JoyceNL10,UllahPSSRRL13} This can lead to enhanced adsorbate-2DEG separation for a fraction of the adsorbates to ZB nanowires, thereby reducing their sensitivity to low levels of adsorbate compared to WZ. At higher adsorbate levels the response should be similar, and we observe that both WZ and ZB come to similar characteristics in the high adsorbate limit for both H$_2$O and EtOH. The difference is likely not quite sufficient for practical gas sensing applications, e.g., in some sort of WZ versus ZB differential arrangement, but has significant bearing on how we should interpret studies of gate performance in InAs nanowires. The obvious conclusion from previous work~\cite{DayehAFM09, DayehSST10, UllahPSSRRL13} is that WZ has generally superior gate performance to ZB, with ZB often giving notably poor performance~\cite{DayehAFM09}. In mixed phase WZ nanowires, the superior performance arises from interface polarization charge effects at axial stacking faults~\cite{DayehAFM09, ChenNL17}. In pure phase WZ nanowires, the performance advantage is entirely due to the greater adsorbate sensitivity of WZ over ZB and the fact that most measurements are obtained in ambient, which is neither too dry nor too wet. In the dry gas limit, we find the gate performance of WZ and ZB to be equally poor. Correspondingly, in the wet gas limit, we find the gate performance to be equally good. Our findings highlight the importance of careful atmospheric control in studies of transport and electronic performance of nanowire transistors. They also make clear the vital role that some form of density-lowering disorder has in improving gate characteristics for InAs nanowire transistors, be it stacking faults in mixed-phase WZ or surface adsorbates on pure-phase nanowires.

\ack This work was funded by the Australian Research Council (ARC) and the University of New South Wales. We thank D.J. Carrad for helpful discussions on experimental aspects of this work. This work was performed in part using the NSW and ACT nodes of the Australian National Fabrication Facility (ANFF).

\end{document}